\documentclass[aps,prl,twocolumn,superscriptaddress,groupedaddress]{revtex4}  

\usepackage{graphicx}  

\usepackage{dcolumn}   

\usepackage{bm}        

\usepackage{amssymb}   

\usepackage{amsmath}

\usepackage{verbatim}

\usepackage[symbol]{footmisc}

\begin{document}









                \title{The role of metallic leads and electronic degeneracies in thermoelectric power generation in quantum dots}

                \author{Achim~Harzheim}

                \author{Jakub~K.~Sowa}

                \author{Jacob~L.~Swett}

                \author{G.~Andrew~D.~Briggs}

                \affiliation{Department of Materials, University of Oxford, Oxford OX1 3PH, United Kingdom}

                \author{Jan~A.~Mol}

                \email{j.mol@qmul.ac.uk}

                \affiliation{Department of Materials, University of Oxford, Oxford OX1 3PH, United Kingdom}

                \affiliation{School of Physics and Astronomy, Queen Mary University of London, London E1 4NS, United Kingdom}

                \author{Pascal~Gehring}

                \email{p.gehring@tudelft.nl}

                \affiliation{Department of Materials, University of Oxford, Oxford OX1 3PH, United Kingdom}

                \affiliation{Kavli Institute of Nanoscience, Delft University of Technology, Delft 2628, Netherlands}

                \date{\today}

\begin{abstract}

The power factor of a thermoelectric device is a measure of the heat-to-energy conversion efficiency in nanoscopic devices. Yet, even as interest in low-dimensional thermoelectric materials has increased, experimental research on what influences the power factor in these systems is scarce.
Here, we present a detailed thermoelectric study of graphene quantum dot devices.
We show that spin-degeneracy of the quantum dot states has a significant impact on the zero-bias conductance of the device and leads to an increase of the power factor.
Conversely, we demonstrate that non-ideal heat exchange within the leads can suppress the power factor near the charge degeneracy point and non-trivially influences its temperature dependence.

\end{abstract}

\maketitle

\begin{figure}[h!]

\includegraphics[width=\linewidth]{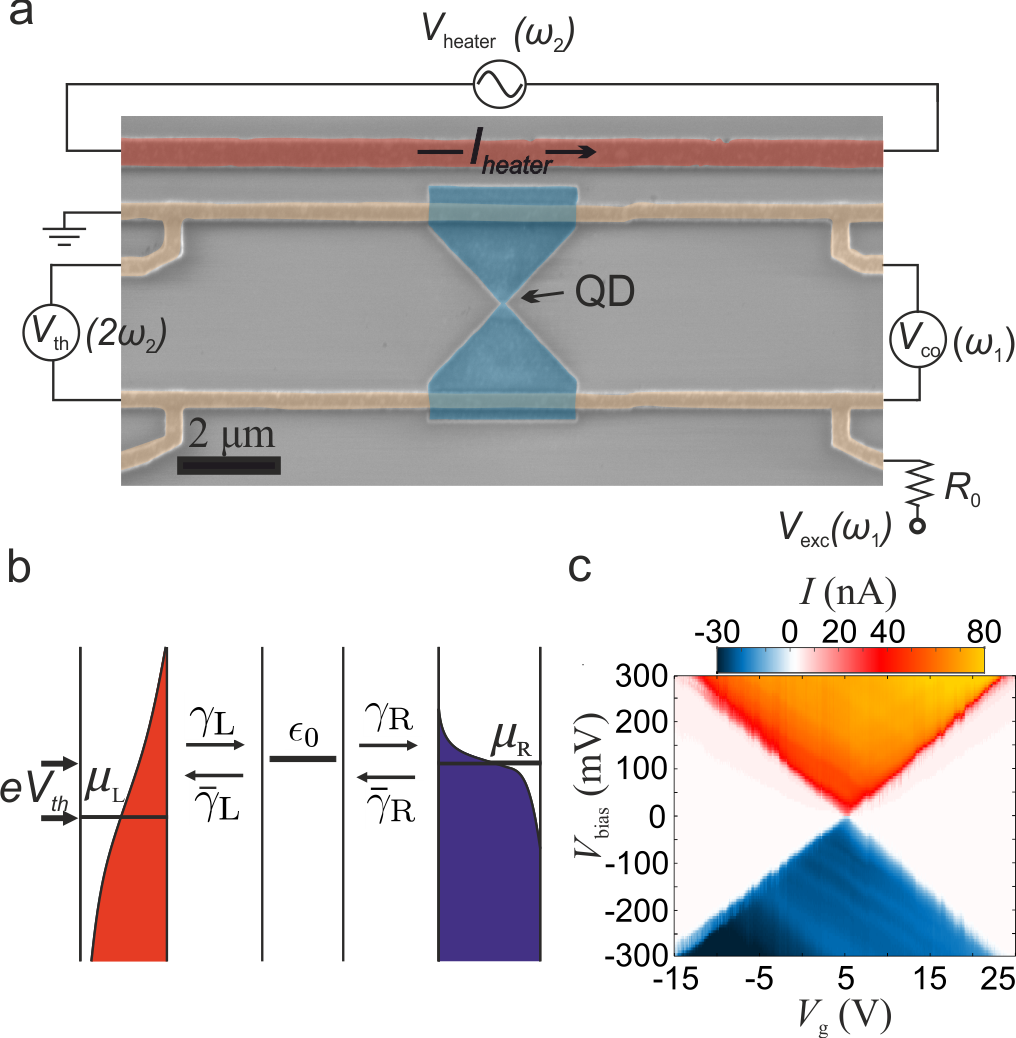}

\caption{(a) False-color SEM image including the measurement schematics. (b) Schematic energy diagram of our device. (c) Stability diagram for device A. }\label{Fig:1:Measurement scheme and stab/cond}

\end{figure}

A thermoelectric device converts a temperature difference between two metallic reservoirs, $\Delta T$, into a thermovoltage, $V_{\mathrm{th}}$. The extent of this conversion is quantified by the Seebeck coefficient (thermopower) defined as: $S = -\frac{\Delta V_{\mathrm{th}}}{\Delta T}$.
However, the Seebeck coefficient alone does not provide a good measure of the heat-to-energy conversion efficiency. In the linear response regime (i.e.~for an operating temperature $T$ such that $\Delta T \ll T$) one should instead consider the power factor: $\mathcal{P} = S^2G$ (where $G$ is the electrical conductance), which is proportional to the maximum power output density of the thermoelectric generator \cite{Liu2016}.
Since the seminal work of Dresselhaus and Hicks \cite{Hicks1993_1D,Hicks1993_0D}, who predicted an increase of the thermoelectric efficiency with decreasing dimensionality, numerous experimental realizations of low-dimensional thermoelectric generators have been achieved \cite{Harman2002,Ibanez2016,Lee2016,Venkatasubramanian2001,Heremans2002}.
Particularly promising here are quantum-dot (QD) devices as they allow for precise control of heat and charge flows.
Recently, the role of the Kondo effect, quantum-interference phenomena and coupling asymmetry in the thermoelectric energy conversion in these systems has been examined theoretically \cite{Sanchez2011,Wierzbicki2011,Krawiec2006,Daroca2018, Gehring2017}. Additionally, recent experimental studies on QD thermoelectric devices demonstrated novel ways of reaching high thermoelectric efficiencies \cite{Thierschmann2015}, even approaching the Carnot efficiency limit \cite{Josefsson2018}. However, the role of intrinsic device features, such as metallic leads, has not yet received much experimental attention. \\
In this work, we experimentally study the thermoelectric properties and power factor of electrostatically-controlled graphene quantum dots (GQD). Our devices comprise a few-nanometer sized GQD located between two graphene leads which are connected to Au leads deposited on a Si/SiO$_2$ chip, see Fig.~\ref{Fig:1:Measurement scheme and stab/cond}a. The GQDs are fabricated using feedback-controlled electroburning of a bow-tie shaped graphene constriction \cite{Gehring2016}, which reliably produces quantum dots with relatively high addition energies \cite{Gehring2017_1}. 
As shown in Fig.~\ref{Fig:1:Measurement scheme and stab/cond}a, to create a temperature gradient across the GQD, we pass an electric current (to induce Joule heating) through a micro-heater located next to the source electrode. We simultaneously measure both the gate dependent conductance, $G$, and thermovoltage, $V_\mathrm{th}$, to avoid artificial offsets between the two quantities (which can lead to wrong estimates of the power factor \cite{Zuev2009}).
To this end, we apply an AC voltage to the source through a resistor in series at a frequency of $\omega_1 = 91$ Hz. The heating current is applied at a frequency $\omega_2 = 17$ Hz while the thermovoltage drop over the quantum dot is measured at the second harmonic, i.e.~at $2\times \omega_2$. The temperature difference between the gold contacts is obtained by a calibration using a four point resistance measurement \cite{Gehring2017}.
When heating up one side of the device, the Fermi distribution of the ``hot" contact gets broader than the one of the ``cold" contact, inducing a thermocurrent between the two leads. In an open circuit configuration, a thermovoltage $V_\mathrm{th}$ will be established so as to nullify the thermocurrent, as schematically shown in Fig.~\ref{Fig:1:Measurement scheme and stab/cond}b.

\textit{Rate-equation model.}---Charge transport through a quantum dot weakly coupled to the metallic leads can be described using a rate-equation (RE) approach. It models the overall transport as a sequence of electron hopping events, assigning a rate for hopping on ($\gamma_l$) and off ($\bar{\gamma}_l$) the QD (where $l = \mathrm{L, R}$ and L (R) stands for the left (right) reservoir), see Fig.~\ref{Fig:1:Measurement scheme and stab/cond}b \cite{Sowa2018}.
In what follows, each lead is found at a temperature $T_l$ which determines the shape of the Fermi distributions: $f_l = [\exp((\epsilon-\mu_l)/k_\mathrm{B}T_l) + 1]^{-1}$, where $k_\mathrm{B}$ is the Boltzmann constant, and $\mu_l$ is the electrochemical potential of the lead $l$.
The quantum-dot energy level (located at $\epsilon_0$) is coupled to the left and right reservoirs with the tunnel coupling strength of $\Gamma_l$. $\epsilon_0$ can be tuned using the gate electrode: $\epsilon_0 = \epsilon_{00} - \alpha \lvert e\rvert V_{\mathrm{g}}$ where $\epsilon_{00}$ is the position of $\epsilon_0$ at zero gate, $\alpha$ is the lever arm and $V_{\mathrm{g}}$ is the gate voltage \cite{Hanson2007}.
We assume that (due to strong electron-electron interactions) only one additional (transport) electron can be found on the QD at any given time.
The electrical current across the QD, $I$, is then given by  \cite{Sowa2019, SI}:
\begin{equation}
    I = \dfrac{e}{\hbar} n_N n_{N+1}\dfrac{\gamma_\mathrm{L} \bar{\gamma}_\mathrm{R}-\gamma_\mathrm{R} \bar{\gamma}_\mathrm{L}}{n_{N+1}(\gamma_\mathrm{L}+\gamma_\mathrm{R})+n_N(\bar{\gamma}_\mathrm{L}+\bar{\gamma}_\mathrm{R})} ~,\label{EQ1:ElectricalcurrentRateEq}
\end{equation}
where $n_N$ and $n_{N+1}$ are the degeneracies of the electronic ground states of $N$ and $N+1$ charge states, respectively. We assume that in the considered QDs only the spin degeneracy plays a role, meaning that $n_N = 1$ and $n_{N+1} = 2$, or \textit{vice versa}.
As we will demonstrate, these factors can be extracted from the temperature-dependent shift of the conductance peak \cite{Beenakker1991}. \\
Similarly, the heat current through the dot, $\dot{Q}$, is \cite{SI}:
\begin{equation}
\dot{Q} = \dfrac{n_N n_{N+1}}{\hbar}\dfrac{\zeta_\mathrm{L}\bar{\gamma}_\mathrm{R}- \bar{\zeta}_\mathrm{L}\gamma_\mathrm{R}}{n_{N+1}(\gamma_\mathrm{L}+\gamma_\mathrm{R})+n_N(\bar{\gamma}_\mathrm{L}+\bar{\gamma}_\mathrm{R})} ~, \label{EQ2:HeatcurrentRateEq}
\end{equation}
where the charge and energy transfer rates in Eqs.~\eqref{EQ1:ElectricalcurrentRateEq} and \eqref{EQ2:HeatcurrentRateEq} are given by:
\begin{align}
 \overset{\scriptscriptstyle(-)}{\gamma}_l  &= 2\: \Gamma_l \int \frac{\mathrm{d} \epsilon}{2 \pi} f_{\pm l} (\epsilon) K(\epsilon)~;\\
 \overset{\scriptscriptstyle(-)}{\zeta}_l  &= 2\: \Gamma_l \int \frac{\mathrm{d} \epsilon}{2 \pi} f_{\pm l} \: (\epsilon) \epsilon \: K(\epsilon)  ~.\label{EQ3:Hopping rates}
\end{align}
In the above equations, lifetime broadening is introduced via the density of states of the QD, which has a Lorentzian line shape: $K(\epsilon) =  \Gamma/[\Gamma^2+(\epsilon-\epsilon_0)^2]$ where $\Gamma = (\Gamma_\mathrm{L}+\Gamma_\mathrm{R})/2$. The Fermi-Dirac distributions are $f_{+ l}(\epsilon) \equiv f_{l}(\epsilon)$ and $f_{- l}(\epsilon) \equiv 1 - f_{l}(\epsilon)$.

\begin{figure}

\includegraphics[width=\linewidth]{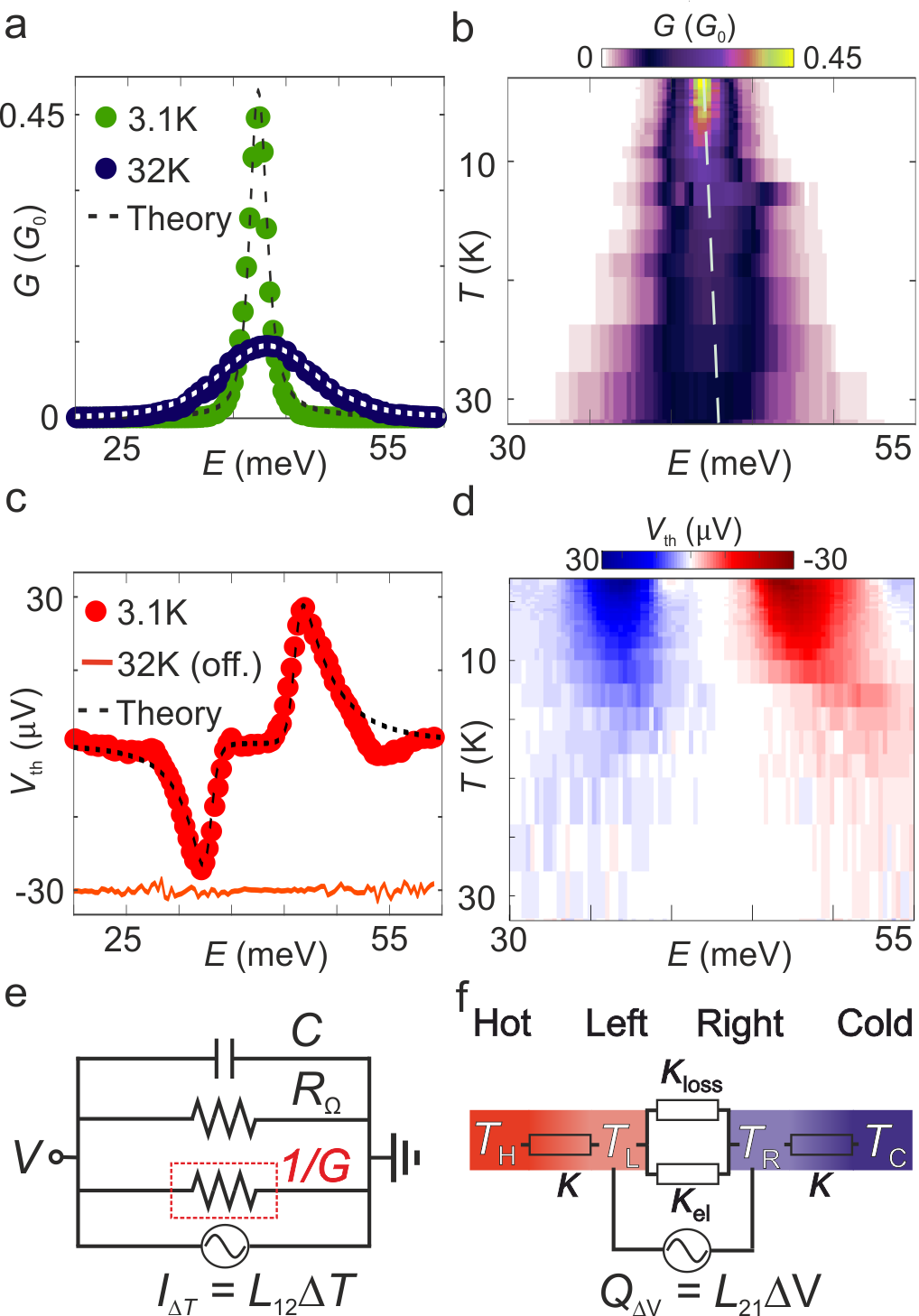}

\caption{(a) Conductance at 3.1K and 32K as a function of $\alpha \lvert e\rvert V_\mathrm{g}$ and the corresponding fits. (b) Conductance as a function of $T$ and $\alpha \lvert e\rvert V_\mathrm{g}$. The dashed line indicates the calculated position of the conductance peaks. (c) Thermovoltage at 3.1K and 32K as a function of $\alpha \lvert e\rvert V_\mathrm{g}$ and the corresponding fit. (d) Thermovoltage as a function of $\alpha \lvert e\rvert V_\mathrm{g}$ and $T$. (e) Schematic of the electrical circuit driven by the thermocurrent $L_{12} \Delta T$, the red dotted line marks the device. (f) Corresponding thermal circuit diagram for our system driven by the heat current $L_{21} \Delta V$. The temperature difference for all  calculations: $\Delta T = 0.52$K.}

\label{Fig:2:Tdependent measurements and Fits}

\end{figure}

\textit{Electronic conductance.}---Fig.~\ref{Fig:1:Measurement scheme and stab/cond}c shows the stability diagram of device A from which we extract the lever arm of $\alpha = 7.9$ meV.
We next consider the zero-bias differential conductance $G = \frac{\mathrm{d}I}{\mathrm{d}V}\Bigr|_{V=0}$ as a function of $V_{\mathrm{g}}$.
We begin by fitting the experimental conductance trace at $T = 3.1$K using the rate-equation model described above, see Fig.~\ref{Fig:2:Tdependent measurements and Fits}a. 
The conductance trace has an approximately Lorentzian lineshape, in agreement with our model and as expected for a single-electron transistor \cite{Kouwenhoven1991}.
From the fit (black dashed line in Fig.~\ref{Fig:2:Tdependent measurements and Fits}a), we obtain tunnel couplings of $\Gamma_\mathrm{L} \approx 0.2$ meV and $\Gamma_\mathrm{R} \approx 1.1$ meV.

Fig.~\ref{Fig:2:Tdependent measurements and Fits}b shows the zero-bias conductance as we increase the temperature from 3.1K to 32K.
Two separate effects can be observed. First, the conductance peak thermally broadens and its maximum decreases accordingly.
Secondly, we observe that the position of the conductance peak shifts with temperature.
This effect has been theoretically predicted by Beenakker \cite{Beenakker1991} and can be understood as originating due to the changes in entropy of the system, as has been discussed elsewhere \cite{Hartman2018}.
The semi-classical theory of Ref.~\cite{Beenakker1991} predicts that (for a doubly degenerate energy level as considered here) the position of the peak should shift by: $\Delta E_\mathrm{shift}(G_{\mathrm{max}}) \approx \mathrm{ln}(2)k_\mathrm{B}T/2$, so that $\mathrm{ln}(2)k_\mathrm{B}/2$ can be regarded as the entropy associated with populating a doubly-degenerate electronic level.
This is also in agreement with our theory.
The expected position of the conductance peak is shown by the white dashed line in Fig.~\ref{Fig:2:Tdependent measurements and Fits}b, which has a slope of $\mathrm{ln}(2)k_\mathrm{B}T/2$.

In Fig.~\ref{Fig:2:Tdependent measurements and Fits}a we further plot the experimental (navy dots) and theoretical (white dashed line) values of the zero-bias conductance at $T=32$K.
Both of the effects discussed above are captured well by our theoretical model (using only the parameters extracted from the earlier fit at 3.1K).
The excellent match between the experimental and theoretical values of electronic conductance also at higher temperature further supports the validity of our theoretical model. 

\textit{Thermovoltage.}---We proceed to analyse the thermovoltage measurements. As discussed, a temperature gradient across the QD is induced using the micro-heater shown in Fig.~\ref{Fig:1:Measurement scheme and stab/cond}a.
$V_\mathrm{th}$ measured at a base temperature of 3.1K is shown in Fig.~\ref{Fig:2:Tdependent measurements and Fits}c.
Similarly to the electrical conductance, the thermovoltage exhibits a gate-dependent behaviour. It switches from negative to positive values as the QD energy level crosses the Fermi energy of the electrodes, indicating a change from a hole to an electron-based thermocurrent \cite{Lunde2005}.
In contrast to what is predicted by the Mott relation (which connects $V_\mathrm{th}$ to the derivative of $G$) \cite{Dzurak1993}, however, we observe a wide region of suppressed $V_{\mathrm{th}}$ around the charge degeneracy point.
As we shall discuss, we attribute this suppression to non-ideal heat exchange within the leads resulting in a different effective temperature difference across the QD, as shown schematically in Fig.~\ref{Fig:2:Tdependent measurements and Fits}f. 
Furthermore, as the temperature increases (from 3.1K to 32K), the amplitude of the thermovoltage signal decreases by an order of magnitude, also in disagreement with a single level model \cite{Gehring2017}.

To explain these effects, we consider the case where the immediate contact regions coupled to the QD are at temperatures $T_\mathrm{L}$ and $T_\mathrm{R}$, respectively, and are not perfectly thermalized with the ``hot" and ``cold" reservoirs (see Fig.~\ref{Fig:2:Tdependent measurements and Fits}e) which are at temperatures $T_\mathrm{H}$ and $T_\mathrm{C}$, respectively.
Instead, the left (right) contact is coupled to the hot (cold) reservoir via a thermal conductance $\kappa(T)$ accounting for both the electrical and phononic heat transport.
Additionally, the thermal conductance through the quantum dot (i.e.~between left and right) consists of an electrical (gate- and temperature-dependent) contribution $\kappa_\mathrm{el}(\epsilon,T)$ and a (strictly temperature-dependent) phonon/substrate contribution $\kappa_{\mathrm{loss}}(T)$. The latter accounts for any phononic heat transport through the substrate as well as the QD.
The resulting thermal circuit diagram is shown in Fig.~\ref{Fig:2:Tdependent measurements and Fits}f. Since it is very challenging to realise perfect thermal contacts between the leads and the QD \cite{Apertet2012}, we believe that similar models should be relevant to most zero-dimensional systems (although the effects discussed here may not alaways be as pronounced as in our study).

To quantify the interplay between the heat and charge flows across the QD and determine the gate-dependent heat flow, we consider the Onsager matrix given by:
\begin{equation}
                \begin{bmatrix}
                I \\ \dot{Q}
                \end{bmatrix}
                =
                \begin{bmatrix}
                L_{11} & L_{12} \\
                L_{21} & L_{22}
                \end{bmatrix}
                \cdot
                \begin{bmatrix}
                \Delta V \\
                \Delta T
                \end{bmatrix}.
\label{EQX:ONSAGER}
\end{equation}
where $L_{ij}$ are the Onsager modes. To obtain $L_{ij}$, we expand $I$ and $\dot{Q}$ to the first order with respect to $\Delta T$ and $\Delta V$. In the linear response regime and assuming small lifetime broadening $(\Gamma \rightarrow 0)$ the modes are given by \cite{SI}:
\begin{equation}
L_{ij} \approx - 2 \frac{e^{2-i}}{\hbar k_{\mathrm{B}}T} \frac{2 \Gamma_\mathrm{L} \Gamma_\mathrm{R}}{(\Gamma_\mathrm{L} + \Gamma_\mathrm{R})} \frac{(\epsilon_0/T)^{j-1}\epsilon_0^{i-1}}{3+3\cosh{(\frac{\epsilon_0}{k_{\mathrm{B}}T_0})}-\sinh{(\frac{\epsilon_0}{k_{\mathrm{B}}T_0})}}
\end{equation}
where we used $n_N = 2$ and $n_{N+1} = 1$.
The magnitude of the electronic heat exchange through the QD (from left to right) is given by $L_{22}$ since the contribution of the $L_{21}$ term can be neglected under open circuit conditions \cite{SI} so that $\kappa_\mathrm{el}= L_{22}$.

Firstly, we note that the phononic thermal transport contribution, $\kappa_\mathrm{loss}$, dominates the thermal transport between the left and right contact off-resonance, i.e.~$\kappa_\mathrm{el} \ll \kappa_\mathrm{loss}$ when $\lvert\epsilon_0\rvert\gg \Gamma, k_\mathrm{B}T$.
Far away from resonance, therefore, the effective temperature gradient across the QD, ($T_\mathrm{L}-T_\mathrm{R}$), is determined only by $\kappa$ and $\kappa_\mathrm{loss}$ \cite{SI}:
\begin{equation}
    T_\mathrm{L}-T_\mathrm{R} = r_\mathrm{\kappa}(T_\mathrm{H}-T_\mathrm{C})~,
\end{equation}
where $r_\mathrm{\kappa} = \kappa/(2\kappa_\mathrm{loss}+\kappa)$. 

On the other hand, accounting for all the heat-flow contributions shown in Fig.~\ref{Fig:2:Tdependent measurements and Fits}f allows us to obtain the effective temperature gradient across the QD as \cite{SI}:
\begin{equation}
    T_\mathrm{L}-T_\mathrm{R} = r_\mathrm{\kappa} \frac{\kappa }{\kappa +2L_{22}+2 L_{21} L_{12} \psi} (T_\mathrm{H}-T_\mathrm{C})~, \label{EQ5:deltaT}
\end{equation}
where $\psi = {R_\mathrm{\Omega}}/{\sqrt{(R_\mathrm{\Omega}+G)^2+(C\omega_2R_\mathrm{\Omega})^2}}$. Here $R_\mathrm{\Omega} = 1 \mathrm{T}\Omega$ is the input impedance of our high-impedance amplifier, and the system capacitance is extracted from our low temperature fit as $C \approx 3$ nF. 
The zero-bias thermocurrent, $I_\mathrm{th}$, is calculated using Eqs.~(\ref{EQ1:ElectricalcurrentRateEq}) and (\ref{EQ5:deltaT}). Accounting for the load resistance ($R_{\Omega}$) and the inherent capacitance of the system ($C$), gives the thermovoltage as \cite{SI,Svensson2012}:
\begin{equation}
   V_{\mathrm{th}} = \frac{I_{\mathrm{th}}R_\mathrm{\Omega}}{\sqrt{(1+R_\mathrm{\Omega}G)^2+(C\omega_2R_\mathrm{\Omega})^2}}~.   \label{EQ6:VthCorrected}
  \end{equation}

As can be inferred from Eqs.~(\ref{EQ5:deltaT}) and (\ref{EQ6:VthCorrected}), the electronic heat flow across the QD, in the presence of non-ideal contacts, should result in a strong suppression of the thermovoltage around the resonance.
This is indeed observed experimentally in Fig.~\ref{Fig:2:Tdependent measurements and Fits}b which also shows the theoretical fit of the rate-equation model (with $C$ and $\Delta T$ as fitting parameters and $r_\mathrm{\kappa} = 1$ held constant). In order to extract $r_\mathrm{\kappa}$ we fit the thermovoltage at corresponding temperatures with $\kappa$ and $\kappa_\mathrm{loss}$ as the fitting parameters and holding all other parameters constant with respect to the low-temperature fit \cite{SI}.\\

\begin{figure}
\includegraphics[width=\linewidth]{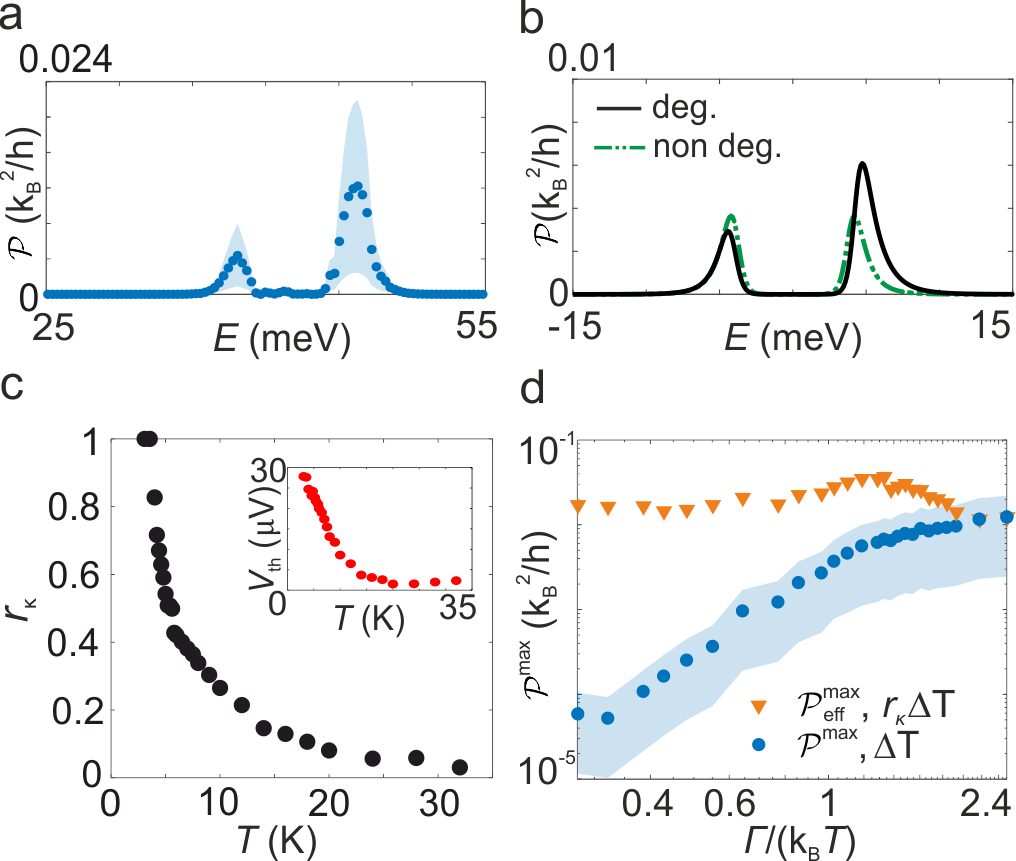}
\caption{(a) Experimental power factor at 3.1K. (b) Calculated power factor for a degenerate (black line) and non-degenerate (dot-dashed green line) energy level, holding all other parameters constant. All as a function of $\alpha \lvert e\rvert V_\mathrm{g}$. (c) Ratio $r_\mathrm{\kappa}$ as a function of $T$. The ratio at each temperature was extracted from fits to the respective thermovoltage \cite{SI} Inset: maximum thermovoltage as a function of temperature. (d) Maximum power factor $\mathcal{P}^\mathrm{max}$ as a function of $T$. The blue dots show $\mathcal{P}^\mathrm{max}$ calculated with the temperature drop between the contacts ($\Delta T$) and the orange triangles $\mathcal{P}_\mathrm{eff}^\mathrm{max}$ calculated with the effective temperature drop $r_\mathrm{\kappa} \Delta T$. In all plots a temperature difference between the two contacts of $\Delta T \approx 0.52$K is used and the propagated error from the temperature calibration is given by the shaded area.}
\label{Fig:3:PF asymmetry and T dependence}
\end{figure}

We next turn to the second non-trivial effect observed in our thermopower measurements: a rapid decrease of $V_\mathrm{th}$ with increasing temperature shown in Figs.~\ref{Fig:2:Tdependent measurements and Fits}c and d and the inset of Fig.~\ref{Fig:3:PF asymmetry and T dependence}c. 
This effect cannot be explained with a simple single-level model (within either the Landauer or a RE approach) and recent experimental findings of an inverse relation studied a metallic island with a continuum of energy levels which is not applicable in our case \cite{Erdman2019}. Nonetheless, we note that thermovoltage decreasing with increasing temperature is almost universally reported for zero-dimensional structures \cite{Staring1993,Molenkamp1994,Small2003,Scheibner2007}. \\
This trend  can be explained by changes in the thermal conductances ($\kappa_\mathrm{loss}$ and $\kappa$) between the reservoirs.
As shown in Fig.~\ref{Fig:3:PF asymmetry and T dependence}c and Eq.~\ref{EQ5:deltaT}, $r_\mathrm{\kappa}$ obtained from our fit and therefore the effective temperature drop over the QD ($T_\mathrm{L}-T_\mathrm{R}$) decreases as the temperature increases. The thermovoltage must therefore follow a similar trend. The behaviour of $r_\mathrm{\kappa}$ could be attributed to a combination of the temperature dependent thermal conductance in graphene which has a non-linear behavior \cite{Seol2010} and size-dependent effects due to the geometry of the device \cite{Xu2014}. We not that the substrate used in this study, SiO$_\mathrm{2}$, has a negligible temperature dependence in our measurement range, therefore $r_\mathrm{\kappa}$ contains mostly information about our nanostructure  \cite{Wingert2016}.\\
Independent of the effects discussed above, the thermovoltage lineshape can be also influenced by the degeneracy of the relevant charge ground states as  reported recently \cite{Kleeorin2019}. Electronic degeneracy can result in a slight asymmetry between the positive and negative peaks of $V_\mathrm{th}$. \\
\textit{Power factor.}---We finally turn to consider the power factor which, as we discussed, can be obtained from the electrical conductance and thermovoltage measurements: $\mathcal{P} =  S^2G = G [V_{\mathrm{th}}/(T_\mathrm{H} - T_\mathrm{C})]^2 $. Fig.~\ref{Fig:3:PF asymmetry and T dependence}a shows the experimental power factor obtained using $G$ and $V_\mathrm{th}$ measured at $T = 3.1$K. Firstly, we note that it exhibits a clear asymmetry with respect to the charge-degeneracy point.
This asymmetry stems from the discussed shift of the conductance peak (since $V_\mathrm{th}$ is always zero at the charge degeneracy point, irrespective of temperature).
Secondly, we observe a strong reduction of the power factor around the charge degeneracy point. This can be attributed to the effects of non-ideal contacts discussed above [which induce a corresponding suppression in $V_\mathrm{th}$, see Eq.~\eqref{EQ5:deltaT}].
Fig.~\ref{Fig:3:PF asymmetry and T dependence}b shows the calculated power factor for the degenerate ($n_N = 2$ and $n_{N+1} = 1$) and non-degenerate ($n_N = n_{N+1} = 1$) QD level, keeping all other parameters constant. 
A good qualitative agreement between experiment and theory can be observed in the case of a degenerate electronic level.\\
Previously, it has been predicted that (for a non-interacting single-level model)  the maximum power factor $\mathcal{P}^\mathrm{max}$ should exhibit a maximum as a function of temperature at $k_\mathrm{B}T  \approx 1.1 \Gamma$ \cite{Gehring2017}.
This is a result of an interplay between $S$ and $G$ which are expected to increase and decrease with increasing $k_\mathrm{B}T/\Gamma$, respectively.
For the doubly-degenerate level considered here, the maximum of the power factor should instead be expected to occur at $k_\mathrm{B}T\approx 1.05 \Gamma$ \cite{SI}.
As we show in Fig.~\ref{Fig:3:PF asymmetry and T dependence}d, however, we find that this trend is not observed experimentally (blue dots) since, in our experiments, the thermovoltage decreases with increasing $T$ (see inset in Fig.~\ref{Fig:3:PF asymmetry and T dependence}c).
A decreasing power factor with increasing temperature was measured in all devices included in this study, even when a less pronounced suppression of the thermovoltage signal was observed \cite{SI}. This suggests that this is a persistent effect which needs to be taken into account when designing a high-efficiency thermoelectric QD generator.\\
However, in our calculation of the power factor we can replace the applied temperature difference ($T_\mathrm{H} - T_\mathrm{C}$) with the effective one, so that the effective power factor is $\mathcal{P}_\mathrm{eff} = G [V_{\mathrm{th}}/(T_\mathrm{L}-T_\mathrm{R})]^2$. We then recover the expected behaviour of the power factor. This is shown in Fig.~\ref{Fig:3:PF asymmetry and T dependence}d (orange triangles) where $\mathcal{P}_\mathrm{eff}^\mathrm{max}$ reaches a maximum for $k_\mathrm{B}T \approx 1.3 \Gamma$, in a relatively good agreement with the theoretical predictions.\\
\textit{Conclusions}.---We investigated the influence of non-ideal contacts and electronic degeneracy on the thermoelectric properties of graphene quantum dots by simultaneously measuring their thermovoltage and conductance. We have shown that the spin degeneracy of the QD level leads to a (temperature-dependent) shift in the electronic conductance peak (as compared to a non-degenerate or non-interacting electronic level). As discussed, this gives rise to an asymmetric enhancement of the thermoelectric power factor.  
Conversely, non-ideal heat exchange within the leads and heat transport through the QD were found to have a deleterious impact on the observed thermovoltage. In particular, due to an efficient electronic heat transport across the quantum dot, this effect results in a strong suppression of $V_\mathrm{th}$ in the vicinity of the charge-degeneracy point. 
Our experimental results are supported by a rate-equation model which successfully captures all the phenomena described above. 
We believe that the effects discussed here are ubiquitous to quantum-dot thermoelectric devices. 
This work, therefore, opens the door to engineering zero-dimensional devices with increased theremoelectric power factor and provides further understanding of phenomena governing heat-to-energy conversion in such systems.\\

This work was supported by the UK EPSRC (Grants EP/K001507/1, EP/J014753/1, EP/H035818/1, EP/K030108/1, EP/J015067/1, and EP/N017188/1). P.G. acknowledges a Marie Skłodowska-Curie Individual Fellowship under Grant TherSpinMol (ID: 748642) from the European Union’s Horizon 2020 research and innovation programme. J.A.M. acknowledges a RAEng Research Fellowship.

\end{document}


	
	
	\title{Supplementary Material:\\ The role of metallic leads and electronic degeneracies in thermoelectric power generation in quantum dots}
	
	\author{Achim~Harzheim}
	\author{Jakub~K.~Sowa}
	\author{Jacob~L.~Swett}
	\author{G.~Andrew~D.~Briggs}
	\affiliation{Department of Materials, University of Oxford, Oxford OX1 3PH, United Kingdom}
	\author{Jan~A.~Mol}
	\email{j.mol@qmul.ac.uk}
	\affiliation{Department of Materials, University of Oxford, Oxford OX1 3PH, United Kingdom}
	\affiliation{School of Physics and Astronomy, Queen Mary University of London, London E1 4NS, United Kingdom}
	\author{Pascal~Gehring}
	\email{p.gehring@tudelft.nl}
	\affiliation{Department of Materials, University of Oxford, Oxford OX1 3PH, United Kingdom}
	\affiliation{Kavli Institute of Nanoscience, Delft University of Technology, Delft 2628, Netherlands}

	\date{\today}

	\maketitle
	\renewcommand{\theequation}{S\arabic{equation}}	

\section{Derivation of $I$ and $\dot{Q}$}

Our graphene quantum dot can be described by the Hamiltonian
\begin{equation}
    H_S = \sum_{\sigma = \uparrow,\downarrow} \epsilon_0 \:a_{\sigma}^\dagger a_{\sigma} + U n_{\uparrow}n_{\downarrow}~,
    \label{SEQ:Hamiltonian}
\end{equation}
where $\epsilon_0$ is the position of the Quantum Dot (QD) energy level with the creation (annihilation) operator $a_{\sigma}^\dagger$ ($a_\sigma$) and $\sigma$ denotes the electron spin. The second term describes the on-site electron-electron repulsion with the strength $U$.
We assume that the double population of the QD is not possible and set $U \rightarrow \infty$.
Using a quantum-master-equation approach \cite{Sowa2018}, we can then write the occupation probability of our QD in terms of hopping on/off rates (where we ignore the repulsive $U n_{\uparrow}n_{\downarrow}$ term in the equations of motion, see Ref.~\cite{Sowa2018}):
\begin{align}
	& P_N =\frac{n_{N+1}(\bar{\gamma}_L+\bar{\gamma}_R)}{n_{N+1}(\gamma_\mathrm{L}+\gamma_\mathrm{R})+n_N(\bar{\gamma}_\mathrm{L}+\bar{\gamma}_\mathrm{R})} ~;\\ 
	& P_{N+1} =\frac{n_N(\gamma_L+\gamma_R)}{n_{N+1}(\gamma_\mathrm{L}+\gamma_\mathrm{R})+n_N(\bar{\gamma}_\mathrm{L}+\bar{\gamma}_\mathrm{R})}~. 
\end{align}\\
Here, $P_N$ ($P_{N+1}$) is the probability of the QD being in the $N$ ($N+1$) charge state. As discussed in the main body of this work, $n_N$ and $n_{N+1}$ are the degeneracies of the electronic ground states of the $N$ and $N+1$ charge states, respectively. The hopping rates are given by Eqs. (3) and (4) in the main body of this work:
\begin{equation}
 \overset{\scriptscriptstyle(-)}{\gamma}_l  = 2\: \Gamma_l \int \frac{\mathrm{d} \epsilon}{2 \pi} f_{\pm l} (\epsilon) K(\epsilon) ~,
\end{equation}
where again lifetime broadening is introduced via the density of states of the QD, which we assume to have a Lorentzian line shape: $K(\epsilon_0) =  \Gamma/[\Gamma^2+(\epsilon - \epsilon_0)^2]$ with $\Gamma = (\Gamma_\mathrm{L}+\Gamma_\mathrm{R})/2$. The Fermi-Dirac distributions are denoted as $f_{+ l}(\epsilon) \equiv f_{l}(\epsilon)$ and $f_{- l}(\epsilon) \equiv 1 - f_{l}(\epsilon)$.\\
Then, the current $I$ can be found at, for instance, the left interface as:
\begin{equation}
	I = \frac{e}{\hbar} [P_N\gamma_{L}-P_{N+1} \bar{\gamma}_{L}].
	\label{SEQCurrent}
\end{equation}
Simplifying Eq.~\ref{SEQCurrent} gives:
\begin{equation}
\begin{aligned}
    & I = \frac{e}{\hbar} n_N n_{N+1}\frac{\gamma_{L}(\bar{\gamma}_L+\bar{\gamma}_R)-\bar{\gamma}_{L}(\gamma_L+\gamma_R)}{n_{N+1}(\gamma_\mathrm{L}+\gamma_\mathrm{R})+n_N(\bar{\gamma}_\mathrm{L}+\bar{\gamma}_\mathrm{R})}\\
    & = \dfrac{e}{\hbar} n_N n_{N+1}\dfrac{\gamma_\mathrm{L} \bar{\gamma}_\mathrm{R}-\gamma_\mathrm{R} \bar{\gamma}_\mathrm{L}}{n_{N+1}(\gamma_\mathrm{L}+\gamma_\mathrm{R})+n_N(\bar{\gamma}_\mathrm{L}+\bar{\gamma}_\mathrm{R})}.
    \label{SEQSimpCurrent}
\end{aligned}
\end{equation}
We can perform a similar operation for the heat current $\dot{Q}$, which can be written in terms of occupation probabilities as
\begin{equation}
    \dot{Q} = \frac{1}{\hbar} [P_N\zeta_{L}-P_{N+1} \bar{\zeta}_{L}]~,
	\label{SEQHCurrent}
\end{equation}
where the energy transfer rates are given by: 
\begin{equation}
     \overset{\scriptscriptstyle(-)}{\zeta}_l  = 2\: \Gamma_l \int \frac{\mathrm{d} \epsilon}{2 \pi} f_{\pm l} \: (\epsilon) \epsilon \: K(\epsilon)  ~.
\end{equation}

\noindent Then,
	\begin{align}
	&\dot{Q} = \frac{1}{\hbar} \left(n_{N+1} \zeta_L P_N - n_N \bar{\zeta}_L P_{N+1} \right)~;\\
	&\dot{Q}= \frac{1}{\hbar}  \frac{n_{N+1}n_N\zeta_L(\bar{\gamma}_L+\bar{\gamma}_R)-n_{N+1}n_N\bar{\zeta}_L(\gamma_L+\gamma_R)}{n_{N+1}\gamma_L+n_{N+1}\gamma_R+n_N\bar{\gamma}_L+n_N\bar{\gamma}_R}~.
	\end{align}
In order to ensure convergence of the above expressions, we take the limit of small lifetime broadening, i.e.~$\Gamma \rightarrow 0$. Then, the term proportional to $(\zeta_L \bar{\gamma}_L - \bar{\zeta}_L \gamma_L)$ vanishes and we can write $\dot{Q}$ as:
\begin{equation}
    \dot{Q} = \dfrac{n_N n_{N+1}}{\hbar} \dfrac{\zeta_\mathrm{L} \bar{\gamma}_\mathrm{R}-\gamma_\mathrm{R} \bar{\zeta}_\mathrm{L}}{n_{N+1}(\gamma_\mathrm{L}+\gamma_\mathrm{R})+n_N(\bar{\gamma}_\mathrm{L}+\bar{\gamma}_\mathrm{R})}~.
    \label{SEQSimHCurrent}
\end{equation}

\section{Linear expansion of $I$ and $\dot{Q}$}

The Onsager reciprocity relation takes the form given by Eq.~(5) in the main body of this work:
\begin{gather}
	\begin{bmatrix}
	I \\ \dot{Q}
	\end{bmatrix}
	=
	\begin{bmatrix}
	L_{11} & L_{12} \\
	L_{21} & L_{22}
	\end{bmatrix}
	\cdot
	\begin{bmatrix}
	\Delta V \\
	\Delta T
	\end{bmatrix}
	\label{SEQOnsager}
\end{gather}
with the current  and heat flow defined in Eqs.~(\ref{SEQSimpCurrent}) and \eqref{SEQSimHCurrent}, respectively. 
In order to obtain the Onsager matrix elements in Eq.~\eqref{SEQOnsager}, we shall expand $I$ and $\dot{Q}$ to the first order with respect to the bias voltage $\Delta V$ and temperature difference $\Delta T$.
First, we note that neglecting lifetime broadening enables us to write the charge transfer rates as 
\begin{equation}
	\gamma_{L/R} = \Gamma_{L/R}  f_{L/R}(\epsilon_0)~,
\end{equation} 
and
\begin{equation}
	\bar{\gamma}_{L/R} = \Gamma_{L/R}  [1-f_{L/R}(\epsilon_0)]~.
\end{equation}
Now, assuming a bias voltage is applied only to the left lead and setting $n_N = 1$, $n_{N+1} = 2$, the expression for current can be simplified to:
\begin{equation}
	 I =\frac{e}{\hbar}\frac{2\Gamma_L \Gamma_R[f_L(\epsilon_0)-f_R(\epsilon_0)]}{\Gamma_L f_L(\epsilon_0)+\Gamma_R f_R(\epsilon_0)+\Gamma_L+\Gamma_R}~.
\end{equation}
To extract the mode $L_{11}$ in Eq.~(\ref{SEQOnsager}). we can expand $I$ linearly as a function of $\Delta V$:
\begin{equation}
	I \approx \frac{e^2\Gamma_L \Gamma_R}{\Gamma_L+\Gamma_R} \frac{1}{k_\mathrm{B} T (3+3\cosh{(\epsilon_0/k_\mathrm{B} T)}-\sinh{(\epsilon_0/k_\mathrm{B} T})} \Delta V~,
\end{equation}
which yields:
\begin{equation}
    L_{11} = \frac{e^2\Gamma_L \Gamma_R}{\Gamma_L+\Gamma_R} \frac{1}{k_\mathrm{B} T (3+3\cosh{(\epsilon_0/k_\mathrm{B} T)}-\sinh{(\epsilon_0/k_\mathrm{B} T}))}~.
\end{equation}
\noindent To obtain $L_{12}$, we expand $I$ to the first order with respect to $\Delta T$. Since
\begin{equation}
\overset{\scriptscriptstyle(-)}{\gamma}_L(T+\Delta T) \approx \overset{\scriptscriptstyle(-)}{\gamma}_L(T) + 2 \Gamma_L \frac{1}{2 \pi} f'(\epsilon_0) [\mp \frac{\epsilon_0}{T}] \Delta T ~,
\end{equation}
we obtain
\begin{equation}
L_{12} = \frac{e\Gamma_L \Gamma_R}{\Gamma_L+\Gamma_R} \frac{\epsilon_0}{(k_\mathrm{B} T)^2 (3+3\cosh{(\epsilon_0/k_\mathrm{B} T)}-\sinh{(\epsilon_0/k_\mathrm{B} T}))}~.
\end{equation}\\
	
\noindent 
We can now again expand the heat flow (to the first order in $\Delta V$ and $\Delta T$) analogously to what we have done for the electric current $I$. We set $\gamma_L \approx \gamma_L^0 + \gamma_L^V \Delta V$ and $\zeta_L \approx \zeta_L^0 + \zeta_L^V \Delta V$ yielding 
\begin{equation}
\dot{Q} \approx \frac{n_{N+1}n_N}{\hbar}\frac{(\zeta_L^0+\zeta_L^V \Delta V)\bar{\gamma}_R - (\bar{\zeta}_L^0+\bar{\zeta}^V \Delta V) \gamma_R}{n_{N+1}\gamma_L^0+n_{N+1}\gamma_L^V \Delta V+n_{N+1}\gamma_R+n_N \bar{\gamma}_L^0+n_N \bar{\gamma}_L^V \Delta V + n_N \bar{\gamma}_R}    
\end{equation}
Then,
\begin{multline}
 L_{21} = \frac{n_{N+1}n_N}{\hbar}\frac{\zeta_L^V\bar{\gamma}_R-\bar{\zeta}_L^V\gamma_R}{n_{N+1} \gamma_L^0+n_{N+1}\gamma_R+n_N\bar{\gamma}_L^0+ n_N\bar{\gamma}_R}\\ - \frac{n_{N+1}n_N(\zeta_L^0\bar{\gamma}_R-\bar{\zeta}_L^0\gamma_R)(n_{N+1}\gamma_L^V+n_N\bar{\gamma}_L^V)}{\hbar(n_{N+1} \gamma_L^0+n_{N+1}\gamma_R+n_N\bar{\gamma}_L^0+ n_N\bar{\gamma}_R)^2}~. \label{L211}     
\end{multline}

\noindent The second term on the right-hand side of Eq.~\eqref{L211} vanishes in the limit of $\Delta V \rightarrow 0$, yielding  the $L_{21}$ mode as:
\begin{equation}
	L_{21} = \frac{1}{\hbar}\frac{n_{N+1}n_N(\zeta_L^V\bar{\gamma}_R-\bar{\zeta}_L^V\gamma_R)}{n_{N+1} \gamma_L^0+n_{N+1}\gamma_R+n_N\bar{\gamma}_L^0+ n_N\bar{\gamma}_R} ~.
\end{equation}

\noindent Using the expressions for the energy transfer rates
\begin{equation}
	\overset{\scriptscriptstyle(-)}{\zeta}_L^V = 2 \Gamma_L  \frac{1}{2 \pi} f'(\epsilon_0) [\mp \epsilon_0]~,
\end{equation}
and taking $n_N = 1$ and $n_{N+1} = 2$, we can simplify $L_{21}$ to:
\begin{equation}
	L_{21} = \frac{e\Gamma_L \Gamma_R}{\Gamma_L+\Gamma_R} \frac{\epsilon_0}{(k_\mathrm{B} T) (3+3\cosh{(\epsilon_0/k_\mathrm{B} T)}-\sinh{(\epsilon_0/k_\mathrm{B} T}))}~.
\end{equation}
	
\noindent Finally, an analogous expansion can be performed to the first order in  $\Delta T$, where 
\begin{equation}
	\overset{\scriptscriptstyle(-)}{\zeta}_L^{\Delta T} = 2 \Gamma_L  \frac{1}{2 \pi} f'(\epsilon_0) [\mp \frac{\epsilon_0}{T}] \epsilon_0~.
\end{equation}
Proceeding analogously to what we have done above, we obtain:
\begin{equation}
	L_{22} = \frac{e\Gamma_L \Gamma_R}{\Gamma_L+\Gamma_R} \frac{\epsilon_0^2}{(k_\mathrm{B} T)^2 (3+3\cosh{(\epsilon_0/k_\mathrm{B} T)}-\sinh{(\epsilon_0/k_\mathrm{B} T}))}~.
\end{equation}

\section{Effective temperature gradient}

In Eq. (7) the effective temperature gradient across the device far away from resonance is given by
\begin{equation*}
    T_L-T_R = r_\kappa (T_H-T_C)~.
\end{equation*}
This expression can be derived from the thermal circuit diagram shown in Fig. 2f. The heat flow between the hot and cold reservoir is given by the thermal conductances between the reservoirs:
\begin{equation}\label{htflo}
    \dot{Q}_{HC} = \frac{1}{\frac{1}{\kappa}+\frac{1}{\kappa}+\frac{1}{\kappa_{loss}}}(T_H-T_C)~.
\end{equation}
The heatflow given in Eq.~\eqref{htflo} must be equal to the heatflow through the QD (between the left and right lead). Since far away from resonance $\kappa_{el} \ll \kappa_{loss}$, the heatflow through the QD is given by:
\begin{equation}
    \dot{Q}_{QD} = \kappa_{loss}(T_L-T_R)~.
\end{equation}
Setting $\dot{Q}_{HC} = \dot{Q}_{QD}$ yields:
\begin{equation}
    T_L-T_R = \frac{\kappa}{2\kappa_{loss}+\kappa} (T_H-T_C) = r_\kappa (T_H-T_C)~.
\end{equation}
As expected, if $\kappa_{loss} \gg \kappa$, then $(T_L-T_R) \rightarrow 0$ since no temperature gradient can build up due to an efficient heat exchange through the QD. Similarly, for $\kappa_{loss} \ll \kappa$, the temperature difference $(T_L-T_R) \rightarrow (T_H-T_C)$ since no parasitic heatflow through the QD takes place.

Closer to the resonance, the second source of parasitic heat flow, $\kappa_{el}$, also has to be taken into account. In order to calculate the temperature difference $(T_L-T_R) $, we again employ Kirchhoff's law on the thermal circuit shown in Fig.~2f which gives the equations for the heat currents going into/leaving the two nodes at temperatures $T_L$ and $T_R$:
\begin{equation}
L_{21} \Delta V+\kappa(T_H-T_L)+( \kappa_{el}+\kappa_{loss} ) (T_R-T_L) = 0~;
\label{SEQ_Heat1}
\end{equation}
and
\begin{equation}
-L_{21} \Delta V+\kappa (T_C-T_R) + (\kappa_{el}+\kappa_{loss})(T_L-T_R) = 0.
\label{SEQ_Heat2}
\end{equation}

\noindent By adding Eq.~(\ref{SEQ_Heat1}) to Eq.~(\ref{SEQ_Heat2}) we obtain
\begin{equation*}
\kappa [T_H+T_C-(T_L+T_R)] = 0 ,
\end{equation*}
which just implies that $T_L+T_R = T_H+T_C$. By subtracting Eq.~(\ref{SEQ_Heat2}) from Eq.~(\ref{SEQ_Heat1}) we can derive an expression for the temperature difference that is driving the thermoelectric effect in our setup:
\begin{equation}\label{kirchoff}
2 L_{21} \Delta V +\kappa [(T_H-T_C)-(T_L-T_R)]-2 ( \kappa_e +\kappa_{ph})(T_L - T_R) = 0.
\end{equation}

Here, $\kappa_e$ is equal to the matrix element $L_{22}$ in the Onsager matrix. If we do not operate under open circuit condition (when $\Delta V = 0$), $L_{21}$ contributes to the heat exchange (see Eq.~5). The small AC current applied across the QD to extract its conductance during the thermovoltage measurements is, however, at a considerably higher frequency (91Hz) than the thermovoltage measurement frequency (17Hz). This leads to an averaging (to zero) of the current on the much slower time scale of the thermovoltage measurements.\\ 
$\Delta V$ in Eq.~\eqref{kirchoff} is just the measured thermovoltage, $V_{th} = \frac{I_{th}R_\mathrm{\Omega}}{\sqrt{(1+G R_\mathrm{\Omega})^2+(4 \pi f R_\mathrm{\Omega} C)^2}} = \beta I_{th}$, where $I_{th} \approx L_{12} (T_L - T_R)$.\\
Since we have also derived expressions for $L_{21}$ and $L_{22}$, we can finally obtain $(T_L - T_R)$ from
\begin{equation}
(T_L - T_R)= \frac{\kappa (T_H-T_C) }{\kappa +2(L_{22}+\kappa_{ph})+2 L_{21} L_{12} \beta}.
\end{equation}
$G$ can be written as $G \approx L_{11}$ or derived via complex step derivation. Using $(T_L - T_R)$ we can calculate a thermocurrent $I_\mathrm{th}$ and the thermovoltage $V_\mathrm{th}$.

\section{Conductance measurement}
In this section, we describe how the conductance through the device is measured experimentally.
In what follows, we define $R_D$ as the device resistance (which is the quantity of interest since the conductance is simply given by: $G=1/R_D$). $V$ is the measured voltage drop over the device, $R_0$ is the external resistor used to transform an applied voltage into a current $I$, and finally, $V_0$ is the applied voltage which generates a current via the external resistor $R_0$ (see Fig. 1a main manuscript).\\
Then, 
\begin{align*}
    & \quad \ R =\frac{V}{I}=\frac{V}{V_0/(R_0+R_D)}=R_0+R_D/V~;\\
	& \Leftrightarrow R-\frac{R_DV}{V_0}=\frac{R_0V}{V_0}~; \\
	& \Leftrightarrow R = \frac{R_0V}{V_0}\frac{1}{1-V/V_0}=\frac{R_0V}{V_0-V}~.
\end{align*}
From here, it is relatively straightforward to obtain the conductance $G$ from the measured quantity $V$ (if the applied voltage $V_0$ and the external resistance $R_0$ are known):
\begin{align}
	G = \frac{V_0-V}{R_0 V}=\frac{V_0}{R_0V}-\frac{1}{R_0}=\frac{I}{V}-G_0~.
\end{align}
Note that the measured conductance is independent of the magnitude of the external resistor except for the constant offset introduced by it. It is therefore not necessary to use a resistor such that $R_0 \gg R_\Omega$ to get an accurate relative measurement of the sample conductance. Here, $R_0$ of 1 M$\Omega$ is used (although we note that a smaller value of $R_0$ would also be sufficient).\\
	
\section{Extraction of Thermovoltage}
Fig.~\ref{fS1} shows a schematic of the experimental measurement circuit.
From Fig.~S1, we can infer an expression for the measured thermovoltage as a function of the load resistance $R_\mathrm{\Omega}$, the inherent system capacitance $C$, and the conductance $G$.

\noindent We begin with $V=Q \: C$, which implies that
\begin{equation}
	\dfrac{\mathrm{d}V}{\mathrm{d}t}= \dfrac{\mathrm{d}Q}{\mathrm{d}t}\dfrac{1}{C} = \dfrac{I}{C}~.
\end{equation}
We can write an expression for our circuit including all elements pictured in Fig.~S1.
\renewcommand{\thefigure}{S\arabic{figure}}
\begin{figure}[h!]
		\includegraphics[width=0.5\linewidth]{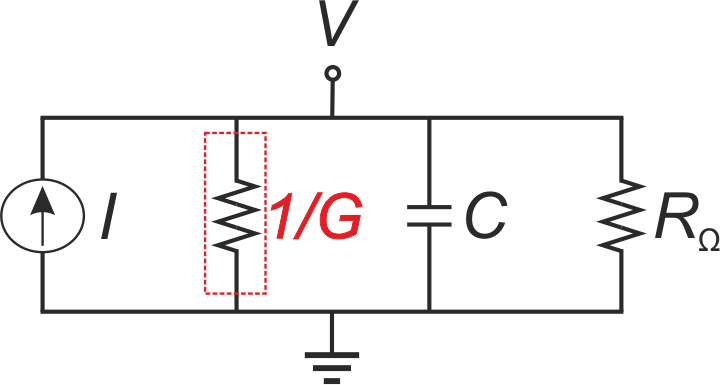}
		\caption{Schematic of the real measurement circuit when probing the thermovoltage $V$, taking into account the capacitance $C$ of our system and the load resistor $R_\mathrm{\Omega}$. The effective device resistance is given by $1/G$.}
		\label{fS1}
\end{figure}
We have
\begin{equation}
I = V (G+1/R_\mathrm{\Omega}) +C \frac{\mathrm{d}V}{\mathrm{d}t}~,
\end{equation}
where both the current $I=I_\mathrm{th} e^{2\mathrm{i}\omega t}$ and the voltage $V=V_\mathrm{th} e^{2\mathrm{i}\omega t}$ are time-dependent. We can then rewrite the equation for the current as
\begin{equation}
I_\mathrm{th} e^{2\mathrm{i}\omega t}=V_\mathrm{th}(G+\frac{1}{R_\mathrm{\Omega}})e^{2\mathrm{i}\omega t}+2V_\mathrm{th}C\mathrm{i}\omega e^{2\mathrm{i}\omega t}~,
\end{equation}
and use it to determine $V_\mathrm{th}$ (which is the quantity of interest): 
\begin{equation}
	V_\mathrm{th} = \frac{I_\mathrm{th}}{G+1/R_\mathrm{\Omega}+2\mathrm{i}\omega C}~.
\end{equation}
Taking the absolute value of $V_\mathrm{th}$ yields
\begin{equation}
	|V_{th}|=\frac{I_{th} R_\mathrm{\Omega}}{\sqrt{(GR_\mathrm{\Omega}+1)^2+4\omega^2R_\mathrm{\Omega}^2C^2}}.
\end{equation}
Where $I_{th}$ is the thermocurrent and the angular frequency is $\omega = 2 \pi f$ so we can write the thermovoltage $V_{th}$ as
\begin{equation}
	V_{th} = \frac{I_{th}R_\mathrm{\Omega}}{\sqrt{(1+G R_\mathrm{\Omega})^2+(4 \pi f R_\mathrm{\Omega} C)^2}}~.
\end{equation}

\section{Temperature Calibration}
\begin{figure}[h!]
		\includegraphics[width=0.5\linewidth]{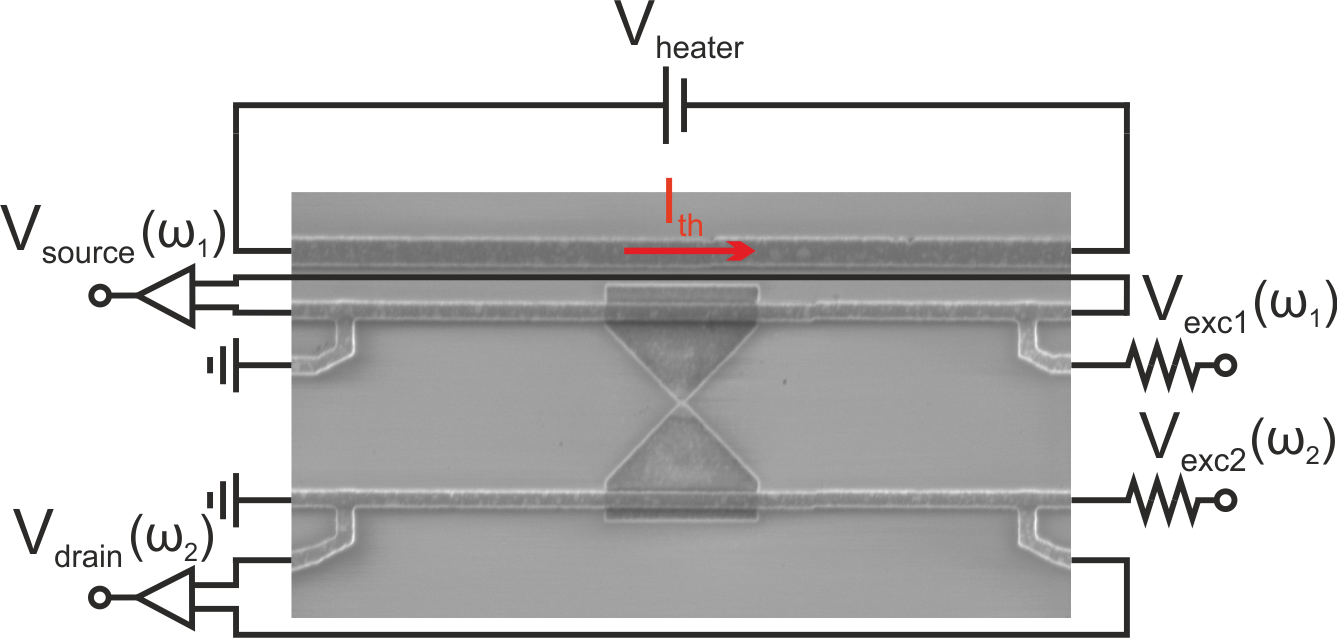}
		\caption{Schematic of the temperature calibration circuit. A voltage $V_\mathrm{heater}$ is applied to the microheater, creating a temperature difference between the source and drain contact. A current is applied to each contact at two different frequencies $\omega_1$ for $V_{exc1}$ and $\omega_2$ for $V_{exc2}$ respectively and the voltage response is measured via a lock-in. Through the measurement of $V_\mathrm{source}$ and $V_\mathrm{drain}$ we can calculate the resistance and calibrate the temperature.}
		\label{fig:S2: Temperature calibration schematic}
\end{figure}

In order to calibrate the temperature difference between the two contacts, an AC method is used as depicted in Fig.~S2. We apply an AC current to each contact separately (the voltage used to source the currents is depicted as $V_\mathrm{exc1}$ and $V_\mathrm{exc2}$ respectively) at frequencies $\omega_1$ and $\omega_2$, which are typically 17 Hz and 91 Hz. At the same time, we measure the voltage response at those frequencies with two seperate lock-ins, which enables us to extract the small resistance changes on the order of 0.3 $\Omega$ of each contact. \\

\begin{figure}[h!]
		\includegraphics[width=0.5\linewidth]{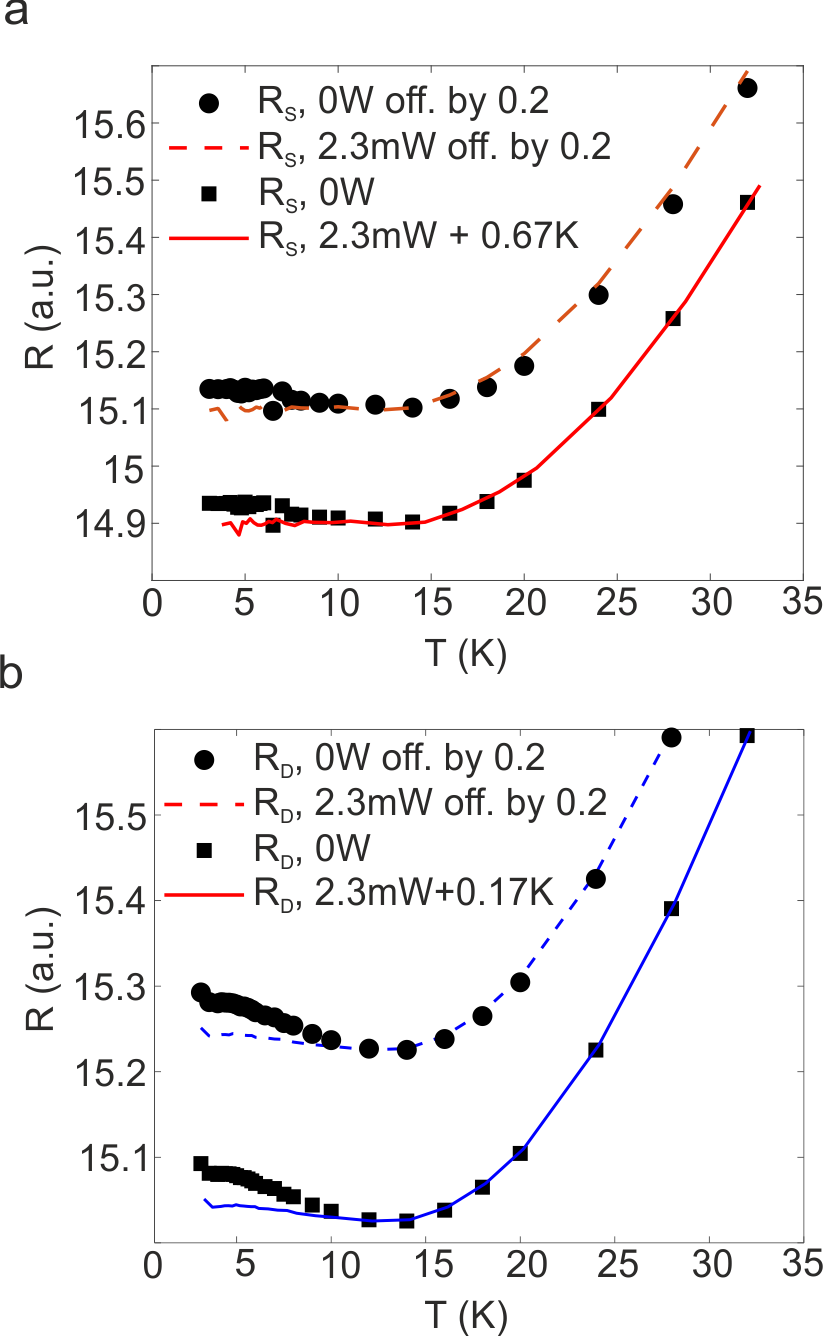}
		\caption{Contact calibration for the source (a) and drain (b) contact. The measured resistance as a funciton of temperature at 0 mW heating power is shown by the dashed line while the resistance at 2.3 mW heating power is symbolized by the dots. Both curves are offset by 0.2 $\Omega$ for clarity. We then plot the 0 mW heating power curve (squares) overlay ed with a shifted 2.3 mW curve, giving us the temperature increase due to Joule heating of the microheater.}
		\label{fig:S2: Temperature calibration schematic}
\end{figure}
	
We measure the resistance of the contacts continuously as we change the temperature and heating power. This enables us to relate a change in resistance to a change in temperature. As can be seen in Fig. S3, for the source (a) and drain (b) contacts, as we change our heating power from 0 to 2.3 mW, the temperature curve shifts. This gives us an indication of how much we heat up the respective contact side. Crucially, however, our measurements are not precise enough to allow for an accurate temperature estimation at low temperatures. 
This is because the resistance of gold does not have a simple linear relation with temperature  below approximately 10 K so that the temperature corresponding to a given resistance value cannot  be easily found at such temperatures. Most importantly, the resistance of the gold heaters has a local minimum at low temperatures resulting in the contact resistance calibration being unreliable at low temperatures. Since the heater power remains constant, we assume a constant temperature difference for the measurements, equal to the higher-temperature calibration value of $(T_H-T_C) \approx 0.5K$ which is also supported by the $(T_H-T_C) =0.52$ extracted from the experimental data. In addition, local fluctuations of the cryostat temperature can add a random error source. Clearly the temperature calibration is the biggest error source in our measurements, and we've assumed an error of $\Delta(T_H-T_C)\approx 20\%$ for all measurements in this study.

As discussed above, the main error source in calculating the power factor comes from the used temperature difference: $T_H-T_C$. In order to calculate the error for the power factor plotted in Fig.~3, we can neglect the measurement error for $G$ and $V_{th}$, as these are much smaller than the uncertainty in determining the temperature difference \cite{Gehring2017}. Additionally, we have also corrected for the effect of the input impedance $R_\omega$  and the intrinsic capacitance in our devices $C$ (see section 5). Using a relative error in  $T_H-T_C$ (and in $T_L-T_R$) of $\approx 20\%$ as determined previously \cite{Gehring2017}, we can propagate the error for the power factor via 
\begin{equation}
    \Delta PF = \sqrt{\left(\frac{\delta PF}{\delta S}\Delta S\right)^2} = 2SG\Delta S = 2 S^2 G \frac{\Delta S}{S}~,
\end{equation}
where $\Delta S$ is the error in $S$.
	
\section{Fits of $V_\mathrm{th}$ at all temperatures}

\begin{figure}[h!]
    \centering
  \includegraphics[width=0.8\linewidth]{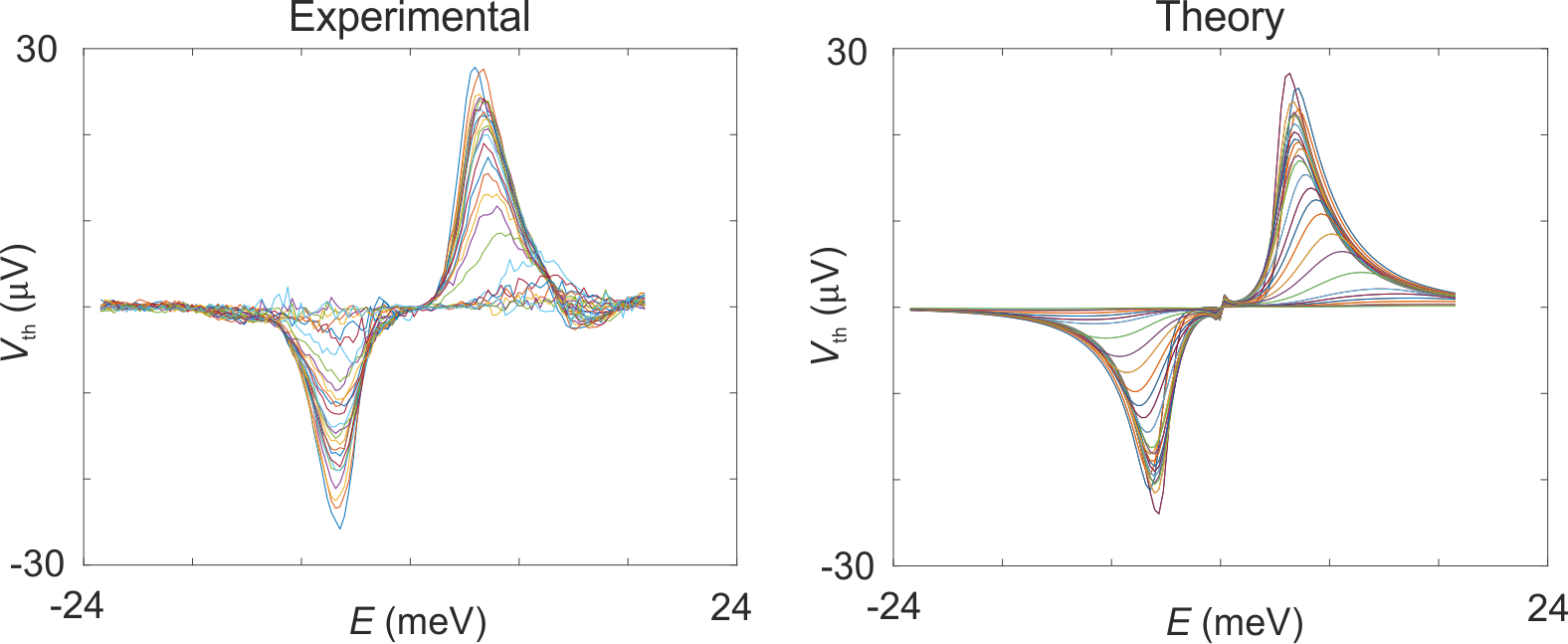}
\captionof{figure}{Experimental thermovoltage as a function of $\alpha \lvert e\rvert V_\mathrm{g}$ (left) and the corresponding fits using our RE theory (right).}
\label{fig:S: Thermovoltage fit}
\end{figure}

In order to extract the ratio between $\kappa_{loss}$ and $\kappa$ that determines the effective temperature drop over the QD, $T_L-T_R = r_\kappa (T_H-T_C)$, we fit the experimental thermovoltage at each measurement point between $3.1$K and $32$K using Eqs. (8) and (9) in the main manuscript. \\ 
Fig.~\ref{fig:S: Thermovoltage fit} shows a good agreement between the experimental measurements and the theoretical fits. The remaining deviations, most noticeably around the charge-degeneracy point, can be explained by the lack of lifetime broadening in our model as well as the high noise level in the experimental data at higher temperatures.

\section{Theoretical prediction of maximum power factor}

\begin{figure}
    \centering
  \includegraphics[width=0.4\linewidth]{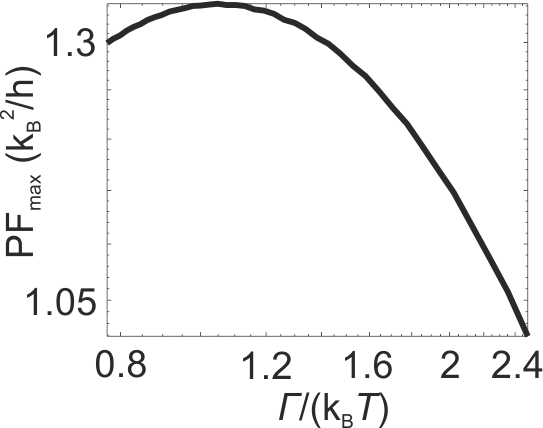}
\captionof{figure}{Theoretical max. power factor as a function of operating temperature and lifetime broadening.}
\label{fig:S: Max PF theory}
\end{figure}

Here, we repeat the calculation of maximum power factor (in the linear response regime, as a function of $\Gamma/k_\mathrm{B}T$) performed in Ref.~\cite{Gehring2017}. In contrast to that work, however, we account for the observed electronic degeneracy [using Eq.~(1) from the main body of this work].
As shown in Fig.~\ref{fig:S: Max PF theory}, similarly as in the case of a non-degenerate level, the maximum power factor exhibits a maximum as a function $\Gamma/k_\mathrm{B}T$ and peaks around $\Gamma \approx 1.05 k_\mathrm{B} T$.
As discussed in the main body of this work, this is in a a relatively good agreement with the experimental data (if the heat-exchange effects within the leads are accounted for).

\section{Devices B-D}

\begin{figure}[h!]
    \centering
  \includegraphics[width=0.8\linewidth]{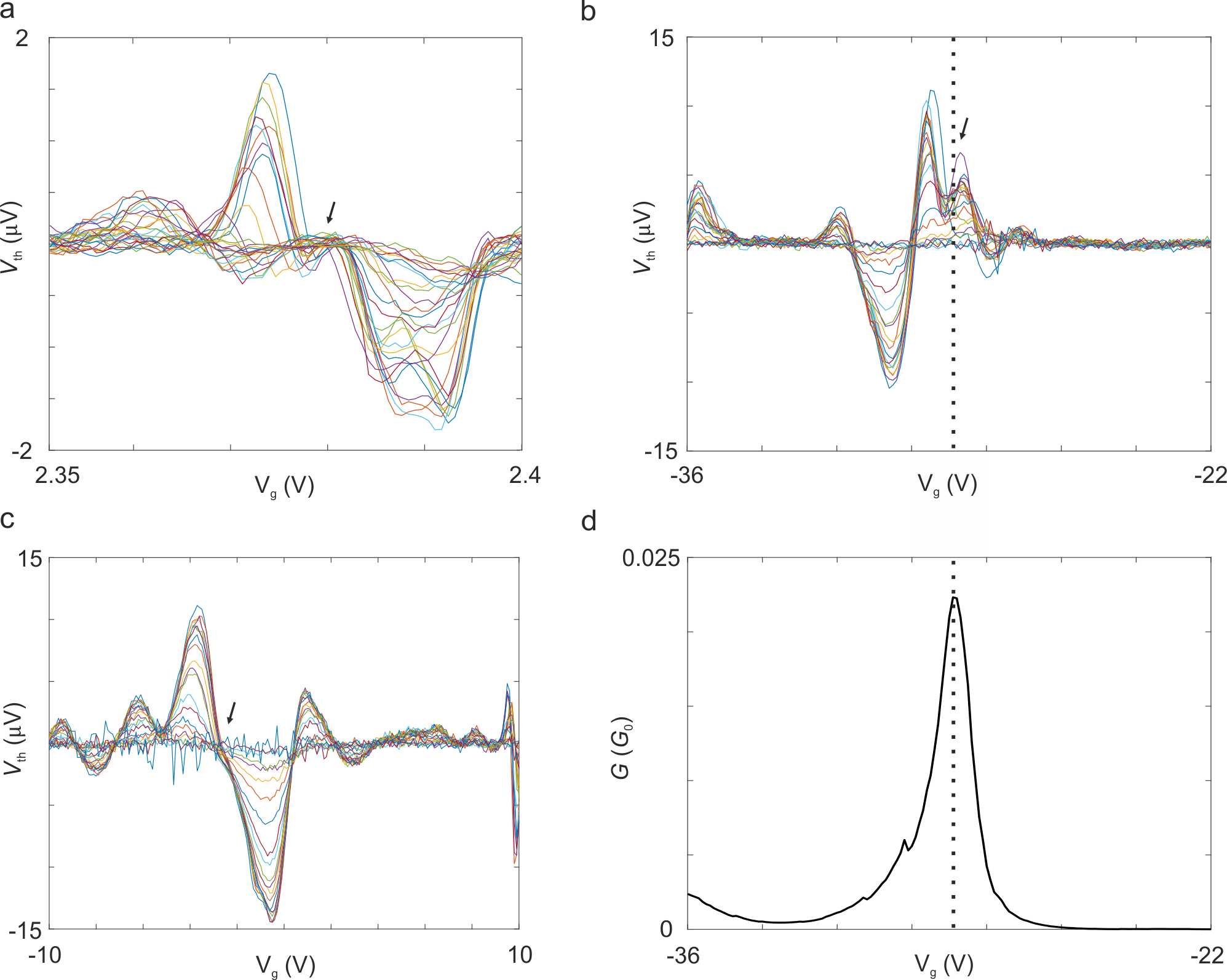}
\captionof{figure}{(a-c) Additional measurements of the gate-dependent thermovoltage for different devices with the operating temperatures ranging from 30 mK to 1800 mK (a) and 4.5K to 32K (b+c). (d) Conductance trace for device C measured at 3.1K and simultaneously to the thermovoltage in (b). The black arrows indicate the position of the conductance peak and therefore the suppression in the thermovoltage traces. The temperature difference is $T_H-T_C \approx 0.5$ K for all measurements.}
\label{fig:S: Thermovoltage OD}
\end{figure}

Here, we present experimental data for the remaining devices were measured in this study. While not all devices show a pronounced suppression around the resonance [for instance device B in Fig.~\ref{fig:S: Thermovoltage OD}a], some degree of suppression can still be observed in the studied devices. In device C [Fig.~\ref{fig:S: Thermovoltage OD}c], a smaller suppression on resonance can be observed which is resulting in a slightly flatter lineshape. 
Interestingly, in device D [Fig.~\ref{fig:S: Thermovoltage OD}b], there is no suppression on resonance visible. 
However, to the right of the resonance a dip in the thermovoltage is observed. From the simultaneously measured conductance, we see that the dip coincides with the conductance peak maximum, indicating that it is the energy dependent heat drop through the QD due to $\kappa_{el}$ that causes the drop in the thermovoltage. \\

The trend of a decreasing power factor with increasing temperature is found in all devices, as can be seen in Fig.~\ref{fig:S: Thermovoltage OD}. This behaviour is observed for a variety of tunnel couplings (from 0.6 meV to 4.1 meV) and is well reproducible. The behaviour of the maximum power factor for high temperatures (left side of the figures) is less reliable as the thermovoltage at this point approaches the noise level and therefore the power factor is prone to errors. In all cases, however, we observe a decrease in the thermovoltage with increasing temperature, which we attribute to the non-ideal heat-exchanging contacts as discussed in the main manuscript.
\\
\begin{center}
  \includegraphics[width=0.8\linewidth]{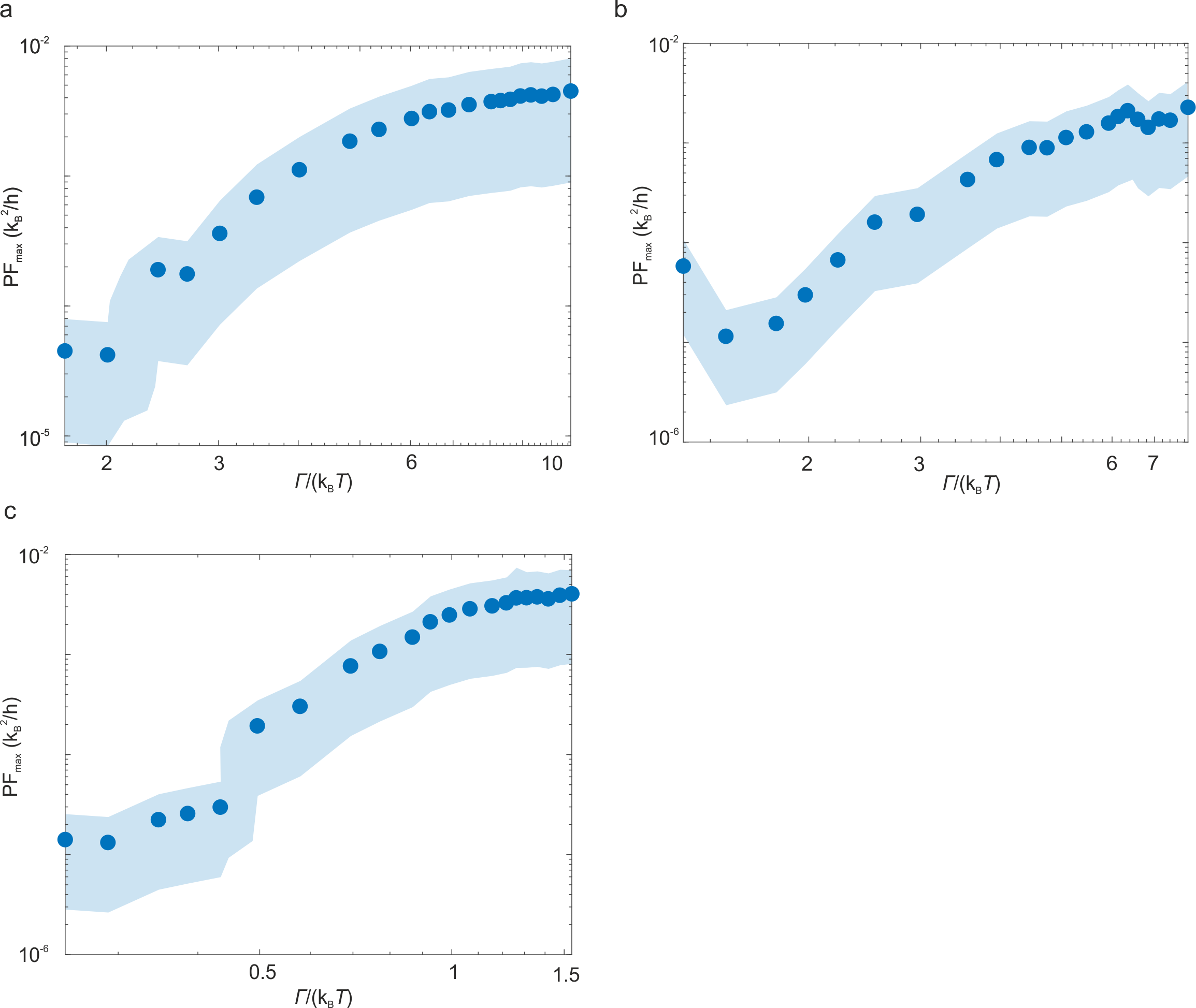}
\captionof{figure}{Additional experimental measurements of the maximum power factor for three different devices between 4.5 K to 31 K with respective coupling strengths of $\kappa \approx 4.1$ meV (s), $\kappa \approx 3$ meV (b) and $\kappa \approx 0.6$ meV (c). The temperature difference is $T_H-T_C \approx 0.5$ K for all measurements.}
\label{fig:S: PF OD}
\end{center}
